\documentclass[%
reprint,
unsortedaddress,
 amsmath,amssymb,
 aps,
]{revtex4-2}

\usepackage[T1]{fontenc} 

\usepackage{mathrsfs}
\usepackage{amsmath}
\usepackage{braket}
\usepackage{dsfont}
\usepackage{hyperref}

\usepackage{slashed} 
\usepackage{enumitem}
\usepackage{physics}
\usepackage{orcidlink}
\usepackage{graphicx}
\usepackage{dcolumn}
\usepackage{bm}
\def\p{\textbf{p}}

\def\0{\textbf{0}}
\def\x{\textbf{x}}
\def\y{\textbf{y}}

\newcommand{\gdualn}[1]{\overset{\:\raisebox{1.5mm}{$\boldsymbol{\neg}$}}{\smash[t]{#1}}}

\newcommand{\widegdual}[1]{\overset{\:\raisebox{1.5mm}{\scalebox{1.4}[1]{$\boldsymbol{\neg}$}}}{\smash[t]{#1}}}

\allowdisplaybreaks
\begin{document}


\title{SUSY meets pseudo-Hermiticity}

\author{Cheng-Yang Lee~\orcidlink{0000-0002-6926-2717}}\email{chengyanglee@outlook.com}\affiliation{Department of Physics and Chongqing Key Laboratory for Strongly Coupled Physics, Chongqing
University, Chongqing 401331, China}

\author{Gloria Cecilia de Le\'on Morales~\orcidlink{0009-0002-1401-4609}}\email{gc.deleonmorales@ugto.mx}
\affiliation{Departamento de F\'isica, Divisi\'on de Ciendias e Ingenier\'ias, Campus Le\'on, Universidad de
  Guanajuato, Loma del Bosque 103, Lomas del Campestre, Le\'on 37150, Guanajuato, Mexico}

\author{Julio Olmos~\orcidlink{0009-0001-1383-2629}}\email{jc.olmosgomez@ugto.mx}
\affiliation{Departamento de F\'isica, Divisi\'on de Ciencias e Ingenier\'ias, Campus Le\'on, Universidad de
  Guanajuato, Loma del Bosque 103, Lomas del Campestre, Le\'on 37150, Guanajuato, Mexico}
  
\author{Carlos A. Vaquera-Araujo~\orcidlink{0000-0001-8578-9263}}\email{vaquera@fisica.ugto.mx}
\affiliation{Secretar\'ia de Ciencia, de Humanidades, Tecnolog\'ia e Innovaci\'on, Insurgentes Sur 1582. Colonia Cr\'edito Constructor, Benito Ju\'arez 03940, Ciudad de M\'exico, Mexico}
\affiliation{Departamento de F\'isica, Divisi\'on de Ciencias e Ingenier\'ias, Campus Le\'on, Universidad de
  Guanajuato, Loma del Bosque 103, Lomas del Campestre, Le\'on 37150, Guanajuato, Mexico}
\affiliation{Dual CP Institute of High Energy Physics, Colima 28045, Colima, Mexico}

\author{Siyi Zhou~\orcidlink{0000-0001-8982-0723}}\email{siyi@cqu.edu.cn}\affiliation{Department of Physics and Chongqing Key Laboratory for Strongly Coupled Physics, Chongqing
University, Chongqing 401331, China}

\date{\today}

\begin{abstract}
In this work, we construct the simplest pseudo-Hermitian quantum field theory that is supersymmetric. This is the pseudo-Hermitian Wess-Zumino model in the sense that it contains a pair of symplectic fermions (anti-commuting scalar fields) that satisfy the Klein-Gordon equation and a spin-half boson that satisfies the Dirac equation. The conventional spin-statistics theorem is circumvented through the use of pseudo-Hermitian conjugation to define field adjoints. To make the supersymmetry manifest, we formulate the pseudo-Hermitian Wess-Zumino model using the superfield formalism. These superfields are Grassmann-odd so it is not possible to construct non-vanishing cubic interactions using only these superfields. We show that this problem can be resolved by coupling the pseudo-Hermitian Wess-Zumino model with the Hermitian Wess-Zumino model while preserving supersymmetry.
\end{abstract}

\maketitle


\section{\label{sec:Intro}Introduction}

The spin-statistics theorem is one of the foundations in quantum field theory (QFT). Within the framework of Lorentz-invariant QFTs with Hermitian Hamiltonians, the theorem states that integer-spin fields must commute while half-integer-spin fields must anti-commute~\cite{Fierz:1939zz,Pauli:1940zz,Feynman:1949hz,PhysRev.110.1450,Pauli1988,Streater:1989vi}. 

In recent years, however, by relaxing the demand of Hermiticity has allowed the exploration of QFTs that circumvent the standard spin-statistics theorem~\cite{LeClair:2007iy,Robinson:2009xm,Ahluwalia:2020jkw,Ahluwalia:2021vfu,Ahluwalia:2022zrm,Bai:2025pvb}. By allowing the Hamiltonians to be pseudo-Hermitian instead of Hermitian, a new class of pseudo-Hermitian QFT where integer-spin fields are anti-commuting while half-integer spin fields are commuting, have been constructed. Within this framework, spin-zero scalar fields are fermionic (symplectic fermions~\cite{LeClair:2007iy,Robinson:2009xm}) while the spin-half spinor fields are bosonic~\cite{Ahluwalia:2020jkw,Ahluwalia:2021vfu,Ahluwalia:2022zrm}.

Supersymmetry, which relates bosonic and fermionic degrees of freedom \cite{Wess:1974tw,Wess:1973kz,Salam:1974yz}, offers a natural arena to further explore pseudo-Hermitian QFT. The simplest $(3+1)$-dimensional supersymmetric Hermitian QFT is the minimal $\mathcal{N}=1$ Wess-Zumino model, which pairs a complex scalar boson with a Majorana fermion. Given the existence of models for symplectic fermions and spin-half bosons, a compelling question arises: \textit{Can we construct a supersymmetric theory that relates symplectic fermions to spin-half bosons?}

In this work, we answer this question in the affirmative by constructing the pseudo-Hermitian Wess-Zumino model. This model contains a pair of symplectic fermions and a spin-half boson. The bosonic and fermionic degrees of freedom are matched by including two anti-commuting scalars to compensate the four degrees of freedom of the spin-half boson, as a Majorana-type construction is not available for commuting spinor fields. We formulate this theory using superspace formalism, and by constructing chiral and antichiral superfields whose component fields exhibit the desired statistics. The resulting free theory is manifestly pseudo-Hermitian and supersymmetric. We further extend our construction to include interactions by coupling the flipped-statistics multiplet to a Hermitian Wess-Zumino multiplet, obtaining an interacting Lagrangian that includes quartic self-interactions for the symplectic fermions as well as novel Yukawa couplings.

\section{Pseudo-Hermitian QFT}\label{sec:ph_qft}

In this section, we briefly review the theory of symplectic fermions and spin-half bosons that are needed to construct the pseudo-Hermitian Wess-Zumino model.

\subsection{Symplectic fermions}

The symplectic fermion is a theory of anti-commuting scalar fields \cite{Bai:2025pvb,LeClair:2007iy} . It consists of a complex scalar field $\phi$ and an adjoint $\gdualn{\phi}$ whose expansions are given by
\begin{align}
    \phi(x)&=\int\frac{d^{3}p}{(2\pi)^{3}}\frac{1}{\sqrt{2E_{\mathbf{p}}}}\left[e^{-ip\cdot x}b(\p)+e^{ip\cdot x}b^{c\dag}(\p)\right],\\
    \gdualn{\phi}(x)&=\int\frac{d^{3}p}{(2\pi)^{3}}\frac{1}{\sqrt{2E_{\mathbf{p}}}}\left[e^{ip\cdot x}b^{\dag}(\p)-e^{-ip\cdot x}b^{c}(\p)\right].
\end{align}
Due to the minus sign in front of $e^{-ip\cdot x}b^{c\dag}$, the fields are fermionic instead of bosonic. The field $\phi$ anti-commutes with its adjoint $\gdualn{\phi}$:
\begin{equation}
    \left\{\phi(t,\x),\gdualn{\phi}(t,\y)\right\}=0,
\end{equation}
with the annihilation and creation operators satisfy the canonical anti-commutation relations
\begin{align}
    \left\{b(\p'),b^{\dag}(\p)\right\}=\left\{b^{c}(\p'),b^{c\dag}(\p)\right\}=(2\pi)^{3}\delta^{3}(\p'-\p).
\end{align}

The field adjoint $\gdualn{\phi}$ is the pseudo-Hermitian conjugate of $\phi$
\begin{equation}
    \gdualn{\phi}(x)=\eta^{-1}_{\phi}\phi^{\dag}(x)\eta_{\phi}
\end{equation}
where $\eta_{\phi}$ is a linear and Hermitian operator. This operator commutes with $b$ and anti-commutes with $b^{c\dag}$. Its solution is given by~\cite{LeClair:2007iy}
\begin{equation}
    \eta_{\phi}=\exp\left[i\pi\int\frac{d^{3}p}{(2\pi)^{3}}b^{c\dag}(\p)b^{c}(\p)\right].
\end{equation}

The symplectic fermions satisfy the Klein-Gordon equation, so their Lagrangian density is given by
\begin{equation}
    \mathcal{L}_{\phi}=(\partial^{\mu}\gdualn{\phi})(\partial_{\mu}\phi)-m^{2}\gdualn{\phi}\phi.
\end{equation}
The conjugate momenta for $\phi$ and $\gdualn{\phi}$ are $\pi=\partial_{t}\gdualn{\phi}$ and $\gdualn{\pi}=-\partial_{t}\phi$, respectively. It is straightforward to show that the fields and their conjugate momenta satisfy the canonical anti-commutation relations. The free Hamiltonian is given by the Legendre transform for $\mathcal{L}_{\phi}$
\begin{align}
    H_{0}(\phi)&=\int d^{3}x\left[\partial_{t}\gdualn{\phi}\partial_{t}\phi+(\nabla\gdualn{\phi})\cdot(\nabla\phi)+m^{2}\gdualn{\phi}\phi\right]\nonumber\\
    &=\int d^{3}p E_{\mathbf{p}}\left[b^{\dag}(\p)b(\p)-b^{c}(\p)b^{c\dag}(\p)\right].\label{eq:H_phi}
\end{align}
Both the Lagrangian density and Hamiltonian are manifestly pseudo-Hermitian
\begin{align}
    \eta^{-1}_{\phi}\mathcal{L}^{\dag}_{\phi}\eta_{\phi}=\mathcal{L}_{\phi},\quad
    \eta^{-1}_{\phi}H^{\dag}_{0}(\phi)\eta_{\phi}=H_{0}(\phi).
\end{align}
From~\eqref{eq:H_phi}, we see that $H_{0}(\phi)$ commutes with $\eta_{\phi}$, so it is simultaneously Hermitian and pseudo-Hermitian
\begin{equation}
    \eta^{-1}_{\phi}H^{\dag}_{0}(\phi)\eta_{\phi}=H^{\dag}_{0}(\phi)=H_{0}(\phi).
\end{equation}

\subsection{Spin-half bosons}\label{sec: Spin_half}

Spin-half bosons originated from the work of Ahluwalia and was later formalized within the framework of pseudo-Hermitian QFT \cite{Bai:2025pvb,Ahluwalia:2022zrm}. Let $\psi$ and $\gdualn{\psi}$ be the massive spinor field and its adjoint for the spin-half boson. Their expansions are given by
\begin{align}
    \psi(x)=\int\frac{d^{3}p}{(2\pi)^{3}}\frac{1}{\sqrt{2E_{\mathbf{p}}}}\sum_{s=\pm1/2}\Big[&e^{-ip\cdot x}u(\p,s)b(\p,s)\nonumber\\
    &+e^{ip\cdot x}v(\p,s)b^{c\dag}(\p,s)\Big],\label{spinor}\\
    \gdualn{\psi}(x)=\int\frac{d^{3}p}{(2\pi)^{3}}\frac{1}{\sqrt{2E_{\mathbf{p}}}}\sum_{s=\pm1/2}\Big[&e^{ip\cdot x}\bar{u}(\p,s)b^{\dag}(\p,s)\nonumber\\
    &-e^{-ip\cdot x}\bar{v}(\p,s)b^{c}(\p,s)\Big]\label{dual_spinor},
\end{align}
where $u$, $v$ are the Dirac spinors and $\bar{u}$, $\bar{v}$ are their corresponding Dirac duals. The minus sign in front of $e^{-ip\cdot x}\bar{v}b^{c}$ makes the spinor fields bosonic
\begin{equation}
    \left[\psi(t,\x),\gdualn{\psi}(t,\y)\right]=0,
\end{equation}
where the annihilation and creation operators satisfy
\begin{align}
    \left[b(\p',s'),b^{\dag}(\p,s)\right]&=\left[b^{c}(\p',s'),b^{c\dag}(\p,s)\right]\nonumber\\
    &=(2\pi)^{3}\delta_{s's}\delta^{3}(\p'-\p).
\end{align}
The adjoint field is defined as 
\begin{equation}
\gdualn{\psi}\equiv\eta^{-1}_\psi\overline{\psi}\eta_{\psi},
\end{equation}
with
\begin{equation}
\eta_\psi=\exp\left[i\pi\int\frac{d^{3}p}{(2\pi)^{3}}\sum_{s=\pm\frac{1}{2}}b^{c\dag}(\p,s)b^{c}(\p,s)\right] .
    \label{eq:eta_psi}
\end{equation}
The bosonic spinor fields satisfy the Dirac equation
\begin{equation}
    \left(i\gamma^{\mu}\partial_{\mu}-m\right)\psi,\label{eq:Dirac_1}
\end{equation}
so the Lagrangian density is
\begin{equation}
    \mathcal{L}_{\psi}=\gdualn{\psi}(i\gamma^{\mu}\partial_{\mu}\psi-m)\psi.\label{eq:L_spinor}
\end{equation}
Direct computation shows that the bosonic spinor fields satisfy the canonical commutation relations and its free Hamiltonian is given by
\begin{align}
    H_{0}(\psi)&=\int d^{3}x\left[-i\gdualn{\psi}i\gamma^{i}\partial_{i}\psi+m\gdualn{\psi}\psi\right]\nonumber\\
    &=\int\frac{d^{3}p}{(2\pi)^{3}}E_{\mathbf{p}}\sum_{s=\pm1/2}\Big[b^{\dag}(\p,s)b(\p,s) \nonumber\\
    &\hspace{3cm}+b^{c}(\p,s)b^{c\dag}(\p,s)\Big].
\end{align}

\section{\label{sec:level1}Pseudo-Hermitian Wess-Zumino model}

The Hermitian Wess-Zumino model consists of a Majorana fermionic field and a complex bosonic scalar field, both satisfying the conventional spin-statistics theorem. The Majorana fermion has two degrees of freedom, which is in one-to-one correspondence with the two bosonic degrees of freedom of the complex scalar field.

In accordance with the new spin-statistics theorem, the corresponding flipped-statistics Wess-Zumino model should be endowed with a global supersymmetry that relates spin-half bosons and symplectic fermions. Note however, there is no bosonic counterpart to Majorana fermions. Spin-half bosons must necessarily have four degrees of freedom on-shell. We cannot reduce its degrees of freedom by identifying the particles with anti-particles. Doing so would make the fields non-local. Our resolution to this problem is construct a Wess Zumino model consisting of \textit{one} spin-half boson and \textit{two} symplectic fermions. This way, the bosonic and fermionic degrees of freedom are in one-to-one correspondence.

In the scalar sector of the multiplet, we have a pair of symplectic fermion fields and their duals
\begin{align}
\phi_i(x)
&=\int\frac{d^{3}p}{(2\pi)^{3}}\frac{1}{\sqrt{2E_{\p}}}\left[e^{-ip\cdot x}b_i(\p)+e^{ip\cdot x}b_i^{c\dag}(\p)\right], \\
\gdualn{\phi}_i(x)
&=\int\frac{d^{3}p}{(2\pi)^{3}}\frac{1}{\sqrt{2E_{\p}}}\left[e^{ip\cdot x}b_i^{\dag}(\p)-e^{-ip\cdot x}b_i^{c}(\p)\right],\label{eq:dual_phi}
\end{align}
with $i=1,2$. These fields are anti-commuting scalars whose dynamics are described by the free massive Lagrangian density 
\begin{equation}
\mathcal{L}_{S}=\sum^{2}_{i=1}\left[(\partial^{\mu}\gdualn{\phi}_i)(\partial_{\mu}\phi_i)-m^{2}\gdualn{\phi}_i\phi_i\right]\,,\label{eq:L_scalar}
\end{equation}
satisfying the Klein-Gordon equation
\begin{equation}(\partial^{\mu}\partial_{\mu}+m^{2})\phi_i=(\partial^{\mu}\partial_{\mu}+m^{2})\gdualn{\phi}_i=0\,.
\end{equation}
The dual fields are defined as
\begin{equation}
\gdualn{\phi}_i=\eta^{-1}_{S}\phi^\dagger_i\eta_{S},
\end{equation}
with the Fock-space operator
\begin{equation}
\eta_{S}=\exp\left[i\pi\int\frac{d^{3}p}{(2\pi)^{3}}\sum^{2}_{i=1}b_i^{c\dag}(\p)b^c_i(\p)\right] ,
    \label{eq:eta}
\end{equation}
that renders the scalar Lagrangian density pseudo-Hermitian 
\begin{equation}
\eta^{-1}_{S}\mathcal{L}_{S}^\dagger\eta_{S}=\mathcal{L}_{S}.    
\end{equation}

In this work, we adopt the Van der Waerden two-component notation following the conventions of \cite{Dreiner:2008tw}. The four-component spin-half field in the chiral basis is written in terms of a pair of left- and right-handed two-component spinors according to
\begin{equation}
    \psi=\begin{pmatrix}
    \xi_\alpha\\
    \chi^\dagger{}^{\dot{\alpha}}
    \end{pmatrix},
\end{equation}
with $\alpha=1,2$ and $\dot{\alpha}=\dot{1},\dot{2}$. The adjoint Dirac field $\overline{\psi}$ and the charge conjugate $\psi^c$ are defined by:
\begin{equation}
\begin{split}    \overline{\psi}&\equiv\psi^\dagger \Pi=\begin{pmatrix}
    \chi^\alpha  &
    \xi^\dagger_{\dot{\alpha}}
    \end{pmatrix},\\
 \psi^{c}&\equiv \mathcal{C}\overline{\psi}^T =  \begin{pmatrix}
    \chi_\alpha  \\
    \xi^\dagger{}^{\dot{\alpha}}
    \end{pmatrix},
\end{split}    
\end{equation}
where the Dirac conjugation matrix $\Pi$ and the charge conjugation matrix $\mathcal{C}$ satisfy
\begin{equation}
\Pi\gamma^\mu \Pi^{-1}=\gamma^{\mu}{}^\dagger,\qquad \mathcal{C}^{-1}\gamma^\mu \mathcal{C}=-\gamma^\mu{}^T.
\end{equation}
In the chiral representation, these matrices can be written in $2\times2$ blocks as
\begin{equation}
\Pi=\begin{pmatrix}
0&\delta^{\dot{\alpha}}{}_{\dot{\beta}}\\
\delta_{\alpha}{}^{\beta} &0
\end{pmatrix}, \qquad 
\mathcal{C}=\begin{pmatrix}
\epsilon_{\alpha\beta}&0\\
0&\epsilon^{\dot{\alpha}\dot{\beta}}
\end{pmatrix},
\end{equation}
together with
\begin{equation}
\gamma^{\mu}=\begin{pmatrix}
0&\sigma^{\mu}_{\alpha\dot{\beta}}\\
\overline{\sigma}^{\mu\dot{\alpha}\beta} &0
\end{pmatrix},\, \sigma^{\mu}\equiv(\mathbf{1},\boldsymbol{\sigma}),\, \overline{\sigma}^{\mu}\equiv(\mathbf{1},-\boldsymbol{\sigma}),
\end{equation}
where $\boldsymbol{\sigma}$ is the three-vector of Pauli matrices. In two-component notation, the relevant fields are
\begin{equation}
\begin{split}
\xi_\alpha(x)&=\int\frac{d^{3}p}{(2\pi)^{3}}\frac{1}{\sqrt{2E_{\mathbf{p}}}}\sum_{s=\pm\frac{1}{2}}
\big[e^{-ip\cdot x}x_\alpha(\p,s)b(\p,s)\\&\qquad\qquad+e^{ip\cdot x}y_\alpha(\p,s)b^{c\dag}(\p,s)\big] ,\\
\chi_\alpha(x)&=\int\frac{d^{3}p}{(2\pi)^{3}}\frac{1}{\sqrt{2E_{\mathbf{p}}}}\sum_{s=\pm\frac{1}{2}}
\big[e^{-ip\cdot x}x_\alpha(\p,s)b^{c\dag}(\p,s)\\&\qquad\qquad+e^{ip\cdot x}y_\alpha(\p,s)b(\p,s)\big] ,\label{eq:chi1}
\end{split}
\end{equation}
with
\begin{equation}
x_\alpha(\p,s)=\sqrt{p\cdot\sigma} \zeta_s, \qquad y_\alpha(\p,s)=2s\sqrt{p\cdot\sigma} \zeta_{-s},    
\end{equation}
and spin states $\zeta_s$, $s=\pm1/2$, specified by a fixed-axis $\mathbf{\hat{s}}$ according to the eigenvalue equation
\begin{equation}
    \frac{1}{2}\boldsymbol{\sigma}\cdot\mathbf{\hat{s}}\zeta_s=s\zeta_s.
\end{equation}
Dotted and undotted indices are related by Hermitian conjugation, and indices are raised and lowered by the Levi-Civita tensors according to 
\begin{equation}
    \xi_\alpha\equiv (\xi_{\dot{\alpha}})^{\dagger},\,\chi_{\dot{\alpha}}\equiv (\chi_\alpha)^{\dagger},\, \xi^{\alpha}\equiv\epsilon^{\alpha\beta}\xi_{\beta}, \, \chi^{\dot{\alpha}}\equiv\epsilon^{\dot{\alpha}\dot{\beta}}\chi_{\dot{\beta}}.
\end{equation}
From these relations, one can easily form invariant bilinear forms in index-free notation
\begin{equation}
    \xi\chi\equiv \xi^\alpha\chi_\alpha, \quad \chi^{\dagger}\xi^\dagger\equiv \chi^{\dagger}_{\dot{\alpha}}\xi^{\dagger\dot{\alpha}}.
\end{equation}
In order to simplify the  two-component notation for the pseudo-Hermitian fields, we adopt the following definition for the spin-half adjoint bosonic spinor:
\begin{equation}
\begin{split}    \gdualn{\psi}=\begin{pmatrix}
    \eta^{-1}\chi^\alpha\eta  &
    \eta^{-1}\xi^\dagger_{\dot{\alpha}}\eta
    \end{pmatrix}\equiv\begin{pmatrix}
   \widetilde{\chi}^\alpha &
   \gdualn{\xi}_{\dot{\alpha}}
    \end{pmatrix},
\end{split}    
\end{equation}
In terms of the above commuting spinors, the Lagrangian density in Eq.(\ref{eq:L_spinor}) for spin-half boson fields reads
\begin{equation}
\begin{split}
\mathcal{L}_\psi=&i\widetilde{\chi}\sigma^\mu\partial_\mu\chi^\dagger+i\gdualn{\xi}\overline{\sigma}^\mu\partial_\mu\xi-m(\widetilde{\chi}\xi+\gdualn{\xi}\chi^\dagger),
\end{split}
\end{equation}
and the Dirac equation in Eq.(\ref{eq:Dirac_1}) splits into the coupled equations
\begin{equation}
i\sigma^\mu\partial_\mu\chi^\dagger-m\xi=0,\qquad i\overline{\sigma}^\mu\partial_\mu\xi-m\chi^\dagger=0.
\end{equation}

The bosonic sector of the multiplet is composed by the fields in Eqs.(\ref{spinor},\ref{dual_spinor}).
In order to simplify the notation for the dual fields, we can define the operator
\begin{equation}
    \eta\equiv\eta_S\eta_\psi,
\end{equation}
that satisfies $\eta^2=1$, and therefore $\eta^{-1}=\eta$. Thus, since $\eta_S$ commutes with $\psi$ and $\eta_\psi$ commutes with $\phi$, we can simply write
\begin{equation}
\gdualn{\phi}=\eta^{-1}\phi^\dagger\eta,\qquad\gdualn{\psi}=\eta^{-1}\overline{\psi}\eta.
\end{equation}

In analogy with the Hermitian Wess-Zumino model, the simplest supersymmetric free theory involving a pair of symplectic fermions and a spin-half boson is described by the following Lagrangian density
\begin{equation}
\mathcal{L}_0=\mathcal{L}_S+\mathcal{L}_\psi,\label{eq:on_shell_free}
\end{equation}
which is pseudo-Hermitian
\begin{equation}
\eta^{-1}\mathcal{L}^\dagger_0\eta= \mathcal{L}_0\,,
\end{equation}
and invariant under the SUSY transformations
\begin{equation}
\begin{split}
\delta\phi_1&=\sqrt{2}\epsilon\xi,\\
\delta\gdualn{\phi}_1&=\sqrt{2}\gdualn{\xi}\epsilon^\dagger,\\
\delta\phi_2&=\sqrt{2}\chi^\dagger\epsilon^{\dagger},\\ \delta\gdualn{\phi}_2&=\sqrt{2}\epsilon\widetilde{\chi},\\
\delta\xi_{\alpha}&=\sqrt{2}m\epsilon_{\alpha} \phi_2-i\sqrt{2}\sigma^\mu_{\alpha\dot{\beta}}\epsilon^{\dagger\dot{\beta}}\partial_\mu\phi_1,\\
\delta\gdualn{\xi}_{\dot{\alpha}}&=\sqrt{2}m\gdualn{\phi}_2\epsilon^{\dagger}_{\dot{\alpha}} +i\sqrt{2}\partial_\mu\gdualn{\phi}_1\epsilon^{\beta}\sigma^{\mu}_{\beta\dot{\alpha}},\\
\delta\widetilde{\chi}_{\alpha}&=\sqrt{2}m\epsilon_{\alpha} \gdualn{\phi}_1-i\sqrt{2}\sigma^\mu_{\alpha\dot{\beta}}\epsilon^{\dagger\dot{\beta}}\partial_\mu\gdualn{\phi}_2,\\
\delta\chi^{\dagger }_{\dot{\alpha}}&=\sqrt{2}m\phi_1\epsilon^{\dagger}_{\dot{\alpha}} +i\sqrt{2}\partial_\mu\phi_2\epsilon^{\beta}\sigma^{\mu}_{\beta\dot{\alpha}}.
\end{split}\label{eq:on_shell_transformations}
\end{equation}
Below, we use the superfield formalism to prove this result.

\section{Superfields}\label{sec:super}
To show that Eq.(\ref{eq:on_shell_free}) is indeed invariant under the rigid supersymmetry transformations in Eq.(\ref{eq:on_shell_transformations}), we formulate the theory in terms of a set of chiral and anti-chiral superfields. In this way, Eq.(\ref{eq:on_shell_free}) emerges from the explicitly supersymmetric $D$-terms of a K\"ahler potential and the $F$-terms of a superpotential. For a superspace with coordinates $x^\mu,\theta^{\alpha},\theta^\dagger_{\dot{\alpha}}$, where the latter are the usual constant complex anti-commuting two-component spinors with mass-dimension $-1/2$, we adopt the usual formulation of $\mathcal{N}=1$ supersymmetry in $3+1$-dimensions with fermionic generators
\begin{eqnarray}
    Q_\alpha&=&-i\left(\frac{\partial}{\partial \theta^\alpha}+i\sigma^\mu_{\alpha\dot{\beta}}\theta^{\dagger\dot{\beta}}\partial_\mu\right),\\
    Q^{\dagger}_{\dot{\alpha}}&=&i\left(\frac{\partial}{\partial \theta^{\dagger\dot{\alpha}}}+i\theta^{\beta}\sigma^{\mu}_{\beta\dot{\alpha}}\partial_\mu\right),
\end{eqnarray}
and supercovariant derivatives
\begin{align}
D_{\alpha}&\equiv\frac{\partial}{\partial \theta^{\alpha}}-i\sigma^{\mu}{}_{\alpha\dot{\beta}}\theta^{\dagger\,\dot{\beta}}\partial_{\mu},\\
D^{\dagger}_{\dot{\alpha}}&\equiv-\frac{\partial}{\partial \theta^{\dagger\dot{\alpha}}}+i\theta^{\beta}\sigma^{\mu}{}_{\beta\dot{\alpha}}\partial_{\mu}.
\end{align} 

To complete the supermutliplet, in order to have the same number of fermionic and bosonic degrees of freedom off-shell, we need to introduce a pair of auxiliary non-dynamic complex scalar fermionic fields $f_i$, $\gdualn{f}_i=\eta^{-1} f^\dagger_i\eta$ ($i=1,2$) of mass dimension two. In terms of the shifted coordinates
\begin{align}
y^\mu&\equiv x^\mu-i\theta\sigma^{\mu}\theta^\dagger,&D^\dagger_{\dot{\alpha}}y^\mu&=0,&D^\dagger_{\dot{\alpha}}\theta&=0 ,\\
y^{\mu\dagger}&\equiv x^\mu+i\theta\sigma^{\mu}\theta^{\dagger},& D_{\alpha}{y}^{\mu\dag}&=0,& D_{\alpha}\theta^\dagger&=0.
\end{align}
We gather these statistics-flipped fields into a collection of anti-commuting scalar chiral and anti-chiral superfields defined as 
\begin{eqnarray}
X&=&\phi_1(y)+\sqrt{2}\theta\xi(y)
+\theta\theta f_1(y),\\
\widegdual{X}&=&\gdualn{\phi}_1(y^\dagger)+\sqrt{2}\gdualn{\xi}(y^\dagger)\theta^\dagger,
+\gdualn{f}_1(y^\dagger)\theta^\dagger\theta^\dagger,\\
Y&=&\phi_2(y^\dagger)+\sqrt{2}\chi^\dagger(y^\dagger)\theta^\dagger
+f_2(y^\dagger)\theta^\dagger\theta^\dagger,\\
\widegdual{Y}&=&\gdualn{\phi}_2(y)+\sqrt{2}\theta\widetilde{\chi}(y)
+\theta\theta \gdualn{f}_2(y).
\end{eqnarray}
They are annihilated by the covariant derivatives $D^\dagger_{\dot{\alpha}}X=D_{\alpha}\widegdual{X}=D^{\dagger}_{\dot{\alpha}}\widegdual{Y}=D_{\alpha}Y=0$, where
\begin{equation}
\widegdual{X}=\eta^{-1}X^\dagger\eta,\, \widegdual{Y}=\eta^{-1}Y^\dagger\eta.
\end{equation}

The transformation of a  superfield $\mathcal{F}(x^\mu,\theta,\theta^\dagger)$ under a rigid SUSY transformation parameterized by two-component anti-commuting spinor parameters $\epsilon^{\alpha},\epsilon^{\dagger}_{\dot{\alpha}}$ is
\begin{equation}
\begin{split}
&\delta\mathcal{F}=i(\epsilon Q+\epsilon^\dagger Q^\dagger)\mathcal{F}\\&=\mathcal{F}(x^\mu-i\theta\sigma^\mu\epsilon^\dagger+i\epsilon\sigma^\mu\theta^\dagger,\theta+\epsilon,\theta^\dagger+\epsilon^\dagger)-\mathcal{F}(x^\mu,\theta,\theta^\dagger).
\end{split}
\end{equation}
For our anti-commuting scalar superfields we have, in components form, the off-shell SUSY transformations,
\begin{equation}
\begin{split}
\delta\phi_1&=\sqrt{2}\epsilon\xi,\\
\delta\gdualn{\phi}_1&=\sqrt{2}\gdualn{\xi}\epsilon^\dagger,\\
\delta\phi_2&=\sqrt{2}\chi^\dagger\epsilon^{\dagger},\\\delta\gdualn{\phi}_2&=\sqrt{2}\epsilon\widetilde{\chi},\\
\delta\xi_{\alpha}&=\sqrt{2}\epsilon_{\alpha} f_1-i\sqrt{2}\sigma^\mu_{\alpha\dot{\beta}}\epsilon^{\dagger\dot{\beta}}\partial_\mu\phi_1,\\
\delta\gdualn{\xi}_{\dot{\alpha}}&=\sqrt{2}\gdualn{f}_1\epsilon^{\dagger}_{\dot{\alpha}} +i\sqrt{2}\partial_\mu\gdualn{\phi}_1\epsilon^{\beta}\sigma^{\mu}_{\beta\dot{\alpha}},\\
\delta\widetilde{\chi}_{\alpha}&=\sqrt{2}\epsilon_{\alpha} \gdualn{f}_2-i\sqrt{2}\sigma^\mu_{\alpha\dot{\beta}}\epsilon^{\dagger\dot{\beta}}\partial_\mu\gdualn{\phi}_2,\\
\delta\chi^{\dagger }_{\dot{\alpha}}&=\sqrt{2}f_2\epsilon^{\dagger}_{\dot{\alpha}} +i\sqrt{2}\partial_\mu\phi_2\epsilon^{\beta}\sigma^{\mu}_{\beta\dot{\alpha}},
\\
\delta f_1&=-i\sqrt{2}\partial_\mu\xi\sigma^\mu\epsilon^\dagger,
\\
\delta \gdualn{f}_1&=i\sqrt{2}\epsilon\sigma^\mu\partial_\mu\gdualn{\xi},\\
\delta f_2&=i\sqrt{2}\epsilon\sigma^\mu\partial_\mu\chi^\dagger,\\
\delta \gdualn{f}_2&=-i\sqrt{2}\partial_\mu\widetilde{\chi}\sigma^\mu\epsilon^\dagger.
\end{split}\label{eq:off_shell_transformations}
\end{equation}
In terms of the superfields, the off-shell Lagrangian of our pseudo-Hermitian Wess-Zumino model reads
\begin{equation}
\begin{split}
\mathcal{L}^{\text{off-shell}}_0=&\left[\widegdual{X}X+\widegdual{Y} Y\right]_D-m\left[\widegdual{X} Y+\widegdual{Y}X\right]_F,
\end{split}
\end{equation}
where the $D$-terms of a general superfield $\mathcal{F}$ are defined as
\begin{equation}
[\mathcal{F}]_D\equiv\int d^2\theta^\dagger d^2\theta \mathcal{F},
\end{equation}
and $F$-terms of a chiral (anti-chiral) superfield $\mathcal{X}$ ($\mathcal{A}$) as
\begin{equation}
[\mathcal{X}]_F\equiv\int d^2\theta \mathcal{X},\qquad [\mathcal{A}]_F\equiv\int d^2\theta^\dagger \mathcal{A}.
\end{equation}
In this way, $\mathcal{L}^{\text{off-shell}}_0$ transforms as a total derivative under Eq.(\ref{eq:off_shell_transformations}).
In terms of the component fields, the off-shell free Lagrangian density is
\begin{equation}\label{eq:off_shell_lag}
\begin{split}
\mathcal{L}^{\text{off-shell}}_0=&\mathcal{L}_\psi+\sum_{i=1}^2\left[(\partial^{\mu}\gdualn{\phi}_i)(\partial_{\mu}\phi_i)+\gdualn{f}_if_i\right]\\
&-m(\gdualn{\phi}_1f_2+\gdualn{\phi}_2f_1+\gdualn{f}_1\phi_2+\gdualn{f}_2\phi_1),
\end{split}
\end{equation}
and the equations of motion for the auxiliary fields are
\begin{equation}\label{eq:eom_f}
    f_1=m\phi_2,\,f_2=m\phi_1,\, \gdualn{f}_1=m\gdualn{\phi}_2,\,\gdualn{f}_2=m\gdualn{\phi}_1,
\end{equation}
showing that the prescription for the duals of the fields $f_i$ is consistent on-shell. Substituting~\eqref{eq:eom_f} into~Eqs.(\ref{eq:off_shell_lag},\ref{eq:off_shell_transformations}), we arrive at the on-shell Lagrangian density in Eq.(\ref{eq:on_shell_free}), and the on-shell SUSY transformations in Eq.(\ref{eq:on_shell_transformations}), while the transformation rules for the auxiliary fields become automatically satisfy the Dirac equation.

\section{Interactions}\label{sec:int}

Since the superfields $X$, $\widegdual{X}$, $Y$ and $\widegdual{Y}$ are Grassmann-odd, it is not possible to form a non vanishing cubic superpotential out of them.
Instead, the simplest interacting theory for our spin-flipped multiplet can be obtained by coupling it with the Wess-Zumino multiplet, packaged into a commuting chiral superfield and its pseudo-Hermitian adjoint
\begin{equation}
\begin{split}
\Phi&=\varphi(y)+\sqrt{2}\theta\lambda(y)+\theta\theta F(y),\\
\widegdual{\Phi}&=\eta^{-1}\Phi^\dagger\eta=\varphi^\dagger(y)+\sqrt{2}\lambda^\dagger(y)\theta^\dagger+\widegdual{F}(y)\theta^\dagger\theta^\dagger
,
\end{split}
\end{equation}
where $\varphi$ is a complex scalar boson, $\lambda$ is a Majorana fermion and $F$ is a complex scalar auxiliary boson. Notice that the physical fields of the conventional Wess-Zumino multiplet $\varphi$ and $\lambda$ commute with the $\eta$ operator. The SUSY transformations of these fields are given by
\begin{equation}
\begin{split}
\delta\varphi&=\sqrt{2}\epsilon\lambda,\\
\delta\varphi^\dagger&=\sqrt{2}\lambda^\dagger\epsilon^\dagger,\\
\delta\lambda_{\alpha}&=\sqrt{2}\epsilon_{\alpha} F-i\sqrt{2}\sigma^\mu_{\alpha\dot{\beta}}\epsilon^{\dagger\dot{\beta}}\partial_\mu\varphi,\\
\delta\lambda^{\dagger }_{\dot{\alpha}}&=\sqrt{2}F^\dagger\epsilon^{\dagger}_{\dot{\alpha}} +i\sqrt{2}\partial_\mu\varphi^\dagger\epsilon^{\beta}\sigma^{\mu}_{\beta\dot{\alpha}},\\
\delta F&=i\sqrt{2}\partial_\mu\lambda\sigma^\mu\epsilon^\dagger,
\\
\delta \widegdual{F}&=-i\sqrt{2}\epsilon\sigma^\mu\partial_\mu\lambda^\dagger.
\end{split}
\end{equation} 
The full pseudo-Hermitian and explicitly supersymmetric interacting Lagrangian is
\begin{equation}
\begin{split}
&\mathcal{L}^{\text{off-shell}}=\left[\widegdual{X} X+\widegdual{Y} Y+\widegdual{\Phi}\Phi\right]_D\\
&\qquad-m\left[\widegdual{X}Y+\widegdual{Y} X\right]_F+\frac{M}{2}\left[\Phi^2+\widegdual{\Phi}^2\right]_F\\
&\qquad+\frac{g}{3!}\left[\Phi^3+\widegdual{\Phi}^3\right]_F+h\left[\widegdual{\Phi} \widegdual{X} Y+\widegdual{Y} X\Phi\right]_F,
\end{split}
\end{equation}
with $M$ as the mass of the fields $\varphi$ and $\lambda$, and dimensionless couplings $g$ and $h$. In terms of the constituting fields, this Lagrangian reads
\begin{equation}
\begin{split}
&\mathcal{L}^{\text{off-shell}}=\mathcal{L}_\psi+\sum_{i=1}^2\left[(\partial^{\mu}\gdualn{\phi}_i)(\partial_{\mu}\phi_i)+\gdualn{f}_if_i\right]\\&\qquad+(\partial^{\mu}\varphi^{\dagger})(\partial_{\mu}\varphi)+\widegdual{F} F+M\left(\varphi F+\varphi^\dagger \widegdual{F}\right)\\&\qquad+i\lambda\sigma^{\mu}\partial_{\mu}\lambda^{\dagger}-\frac{M}{2}\left(\lambda\lambda+\lambda^\dagger\lambda^\dagger\right)\\&\qquad-m(\gdualn{\phi}_1f_2+\gdualn{\phi}_2f_1+\gdualn{f}_1\phi_2+\gdualn{f}_2\phi_1)\\&\qquad+\frac{g}{2}\left[\varphi^2 F+(\varphi^\dagger)^2 \widegdual{F}-\varphi\lambda\lambda-\varphi^\dagger\lambda^\dagger\lambda^\dagger \right]\\&\qquad+h\left[\varphi(\gdualn{\phi}_2 f_1+\gdualn{f}_2 \phi_1)+F\gdualn{\phi}_2\phi_1\right]\\&\qquad+h\left[\varphi^{\dagger}(\gdualn{\phi}_1f_2+\gdualn{f}_1\phi_2)+\widegdual{F}\gdualn{\phi}_1\phi_2\right]
\\&\qquad+h\left[\varphi\widetilde{\chi}\xi+\gdualn{\phi}_2\xi\lambda+\widetilde{\chi}\lambda\phi_1\right]\\&\qquad+h\left[\varphi^{\dagger}\gdualn{\xi}\chi^\dagger+\gdualn{\phi}_1\lambda^\dagger\chi^\dagger+\lambda^\dagger\widegdual{\xi}\phi_2\right].
\end{split}
\end{equation}
The equations of motion for the auxiliary fields are
\begin{equation}
\begin{split}
F&=-M\varphi^\dagger-\frac{g}{2}(\varphi^\dagger)^2-h\gdualn{\phi}_1\phi_2\\
\widegdual{F}&=-M\varphi-\frac{g}{2}\varphi^2-h\gdualn{\phi}_2\phi_1\\
    f_1&=m\phi_2-h\varphi^\dagger\phi_2,\\
    f_2&=m\phi_1-h\varphi\phi_1,\\ \gdualn{f}_1&=m\gdualn{\phi}_2-h\varphi\gdualn{\phi}_2,\\
    \gdualn{f}_2&=m\gdualn{\phi}_1-h\varphi^\dagger\gdualn{\phi}_1,
\end{split}
\end{equation}
Substituting these equations of motion back into $\mathcal{L}$, we obtain the on-shell Lagrangian
\begin{equation}
\mathcal{L}=\mathcal{L}_0+\mathcal{L}_{\varphi}++\mathcal{L}_{\lambda}+\mathcal{L}_\mathrm{I}+\mathcal{L}_\mathrm{Y}.
\end{equation}
Here, $\mathcal{L}_0$ is given by Eq.(\ref{eq:on_shell_free}), $\mathcal{L}_{\varphi}$ is the free Lagrangian of the commuting scalar
\begin{equation}
\begin{split}
\mathcal{L}_{\varphi}=&(\partial^{\mu}\varphi^{\dagger})(\partial_{\mu}\varphi)-M^2\varphi^\dagger\varphi,
\end{split}
\end{equation}
and the free Majorana spinor is described by
\begin{equation}
\begin{split}
\mathcal{L}_{\lambda}=&i\lambda\sigma^{\mu}\partial_{\mu}\lambda^{\dagger}-\frac{M}{2}\left(\lambda\lambda+\lambda^\dagger\lambda^\dagger\right)\\=&\frac{i}{2}\overline{\Lambda}\gamma^\mu\partial_\mu\Lambda-\frac{M}{2}\overline{\Lambda}\Lambda,
\end{split}
\end{equation}
with
\begin{equation}
\Lambda=\Lambda^{c}=\mathcal{C}\overline{\Lambda}^{T}=\begin{pmatrix}
    \lambda_\alpha\\
    \lambda^\dagger{}^{\dot{\alpha}}
    \end{pmatrix}.
\end{equation}
The scalar interaction Lagrangian $\mathcal{L}_\mathrm{I}$ is given by
\begin{equation}
\begin{split}
\mathcal{L}_\mathrm{I}=&-\frac{g^2}{4}(\varphi^\dagger\varphi)^2-\frac{gM}{2}\left[\varphi^\dagger\varphi^2 +(\varphi^\dagger)^2 \varphi\right]\\&+h^2\left[\gdualn{\phi}_1\phi_1\gdualn{\phi}_2\phi_2-\varphi^{\dagger}\varphi\left(\gdualn{\phi}_1\phi_1+\gdualn{\phi}_2\phi_2\right)\right]\\&-\frac{gh}{2}\left[(\varphi^\dagger)^2\gdualn{\phi}_2\phi_1+\varphi^2\gdualn{\phi}_1\phi_2\right]\\&-Mh \left(\varphi^\dagger\gdualn{\phi_2}\phi_1+\varphi\gdualn{\phi_1}\phi_2\right)\\&+mh(\varphi^\dagger+\varphi)(\gdualn{\phi}_1\phi_1+\gdualn{\phi}_2\phi_2),
\end{split}
\end{equation}
and includes the quartic self-interaction for the symplectic fermions analyzed in \cite{Lee:2023aip}. Finally, the Yukawa Lagrangian $\mathcal{L}_\mathrm{Y}$ can be written as
\begin{equation}
\begin{split}
\mathcal{L}_\mathrm{Y}=&-\frac{g}{2}\left[\varphi\lambda\lambda+\varphi^\dagger\lambda^\dagger\lambda^\dagger \right]\\&+h\left[\varphi\widetilde{\chi}\xi+\gdualn{\phi}_2\xi\lambda+\widetilde{\chi}\lambda\phi_1\right]\\&+h\left[\varphi^{\dagger}\gdualn{\xi}\chi^\dagger+\gdualn{\phi}_1\lambda^\dagger\chi^\dagger+\lambda^\dagger\gdualn{\xi}\phi_2\right]\\=&-\frac{g}{2}\left[\varphi\overline{\Lambda}_R\Lambda_L+\varphi^\dagger\overline{\Lambda}_L\Lambda_R \right]\\&+h(\varphi\gdualn{\psi}_R\psi_L+\varphi^\dagger\gdualn{\psi}_L\psi_R)\\&+h(\gdualn{\phi}_1\overline{\Lambda}_L\psi_R+\gdualn{\psi}_R\Lambda_L\phi_1)\\&-h(\gdualn{\phi}_2\overline{\Lambda}_R\psi_L+\gdualn{\psi}_L\Lambda_R\phi_2),
\end{split}
\end{equation}
with 
\begin{equation}
\psi_{L,R}=\frac{1}{2}(1\mp\gamma^5)\psi,
\end{equation}
where $\gamma^5$ is the chiral operator 
\begin{equation}
\gamma^5=i\gamma^0\gamma^1\gamma^2\gamma^3.
\end{equation}

\section{Conclusions}
In this paper,  we have constructed the first consistent supersymmetric theory for QFT with flipped statistics. The departure from the usual spin-statistics connection is achieved by relaxing the requirement of Hermiticity in favor of pseudo-Hermiticity. The multiplet presented here consists of two anticommuting complex scalars known as symplectic fermions, that satisfy the Klein-Gordon equation; two commuting two-component spin-half fields that satisfy the Dirac Equation and furnish Bose-Einstein statistics; and two non-dynamical anticommuting complex scalar auxiliary fields that on-shell become proportional to the symplectic fermions. The resulting model is analogous to the Wess-Zumino model but in its minimal form requires two anticommuting chiral superfields and their duals. The simplest interacting theory  has been studied. This theory couples the flipped-statistics multiplet with a single regular Wess-Zumino multiplet. The resulting superpotential generates, besides the usual interactions for the Wess-Zumino fields, new interactions with the statistics-flipped fields and in particular, a quartic interaction for the symplectic fermions and Yukawa interactions.

This work is the first step towards the study of pseudo-Hermitian supersymmetric QFTs, a novel area of research. Interesting topics to explore include: super-Poincar\'e symmetry and its representations, along the lines studied in \cite{Sablevice:2023odu};  vector superfields and the construction of pseudo-Hermitian gauge theories mirroring \cite{Wess:1974tw,Wess:1973kz,Salam:1974yz}; the study of how some aspects of supersymmetry might be modified with the implementation of pseudo-Hermiticity, as the non-renormalization theorem, and the unitarity of the interacting theory \cite{Lee:2023aip}; the promotion of rigid $\mathcal{N}=1$ transformations to local $\mathcal{N}=1$ SUSY; the study of SUSY breaking mechanisms in this class of theories; and the possibility that a  particle with flipped statistics may become the lightest supersymmetric particle and therefore play the role of a weakly interacting massive particle dark matter candidate. Regarding non-fundamental areas of research, this new class of pseudo-Hermitian supersymmetric models might find applications in areas where SUSY concepts have already been proved useful, like condensed matter physics \cite{Efetov:1997fw} and integrated optics \cite{Miri:2013cdj}.
	
\begin{acknowledgments}
C.A.V.-A. is supported by the Secretar\'ia de Ciencia, Humanidades, Tecnolog\'ia e Innovaci\'on (Secihti) Investigadoras e Investigadores por M\'exico project 749 and SNII 58928. SZ is supported by the Natural Science Foundation of China under Grant No.12347101, No.2024CDJXY022 and No.CSTB2024YCJH-KYXM0070 at Chongqing University.
\end{acknowledgments}

\newpage
\appendix

\bibliography{bibliography}

\end{document}